\documentclass[prb, preprint, letter, superscriptaddress]{revtex4-1}
\usepackage{amssymb}
\usepackage{amsmath}
\usepackage{bbm}
\usepackage{graphicx}
\usepackage{color}

\newcommand{\rif}{\vec{r}_i^{\,f}}
\newcommand{\rjf}{\vec{r}_j^{\,f}}

\newcommand{\aiu}{\alpha_i^{\uparrow}}
\newcommand{\aid}{\alpha_i^{\downarrow}}
\newcommand{\aju}{\alpha_j^{\uparrow}}
\newcommand{\ajd}{\alpha_j^{\downarrow}}

\begin{document}

\author{Shenshen Wang}
\author{Peter G. Wolynes}

\title{Microscopic theory of the glassy dynamics of passive and active network materials}
\date{\today}

\begin{abstract}
Signatures of glassy dynamics have been identified experimentally for a rich variety of materials in which molecular networks provide rigidity. Here we present a theoretical framework to study the glassy behavior of both passive and active network materials. We construct a general microscopic network model that incorporates nonlinear elasticity of individual filaments and steric constraints due to crowding. Based on constructive analogies between structural glass forming liquids and random field Ising magnets implemented using a heterogeneous self-consistent phonon method, our scheme provides a microscopic approach to determine the mismatch surface tension and the configurational entropy, which compete in determining the barrier for structural rearrangements within the random first order transition theory of escape from a local energy minimum.
The influence of crosslinking on the fragility of inorganic network glass formers is recapitulated by the model.
For active network materials, the mapping, which correlates the glassy characteristics to the network architecture and properties of nonequilibrium motor processes, is shown to capture several key experimental observations on the cytoskeleton of living cells: Highly connected tense networks behave as strong glass formers; intense motor action promotes reconfiguration.
The fact that our model assuming a negative motor susceptibility predicts the latter suggests that on average the motorized processes in living cells do resist the imposed mechanical load.
Our calculations also identify a spinodal point where simultaneously the mismatch penalty vanishes and the mechanical stability of amorphous packing disappears.
\end{abstract}

\hyphenation{}

\maketitle

\section{Introduction}

Materials in which molecular networks provide rigidity are ubiquitous, ranging from passive systems such as rubber and silica glasses to active systems, such as the crosslinked polymer networks driven by energy-consuming motor proteins that constitute the cytoskeleton of eukaryotic cells. Both active and passive network materials exhibit features of glassy dynamics in parts of their phase diagrams.
Eukaryotic cells possess a cytoskeleton that stiffens under tension while having an intracellular space crowded with macromolecules and organelles that can resist compression. Recent experiments have established striking similarities between the behavior of a living cytoskeleton and that of inert nonequilibrium soft glasses \cite{cytoskeletal slow dynamics}. Studies of the mechanical and dynamical behavior of osmotically compressed cells further suggest that cells under compressive stress behave as strong glass formers \cite{osmotically compressed cell}. At the same time, F-actin disruption and ATP-dependent nonequilibrium processes have been shown to strongly modulate the glass transition behavior.

Here we develop a framework for studying the glassy behavior of network materials and report calculations for a model of the cytoskeleton to illustrate the concepts. Our scheme is based on the explicit analogy between a structural glass forming liquid and a short range disordered Ising magnet used earlier to study the Lennard-Jones glass \cite{magnetic analogy}. This mapping correlates the glassy characteristics with the network architecture (density and connectivity) and the properties of motor-driven active processes. The approach provides an explicit microscopic route to calculate from the force laws the configurational entropy and the mismatch energy or surface tension, which are competing factors in determining the barrier for structural rearrangements within the random first order transition (RFOT) theory \cite{Xia PNAS}.
Site-dependent Debye-Waller factors, which characterize the extent of localized motion, are determined by a heterogeneous self-consistent phonon (SCP) method.
Our calculations identify a spinodal point at a dynamic transition packing fraction where the mismatch surface tension and mechanical stability simultaneously vanish \cite{Kirkpatrick and Wolynes 1987 PRA, Kirkpatrick and Wolynes 1987 PRB, Kirkpatrick Thirumalai Wolynes 1989 PRA, static correlation length, surface tension, amorphous BC, barrier softening, elasticity_interface}.

Combining the magnetic analogy and self-consistent phonon method allows us to build a correspondence between the theory and several experimental observations: The statistics of the random fields and random interactions calculated from the present network model indicates a possible ideal glass transition underlying the observed glassy behavior of highly connected tense networks made of stiff filaments. When there is a higher degree of connectivity a crossover to strong liquid behavior takes place, both for passive and active materials. Load-resisting motor processes are shown to facilitate structural rearrangements by expanding the accessible configurational state space and lowering the reconfiguration barrier.





\section{A General Microscopic Network Model}

The microscopic model we use can be thought of as an amorphous ``cat's cradle" with excluded volume \cite{CC, TY2, CSK JCP}. It consists of a crosslinked network of nonlinear elastic filaments that stretch elastically with effective stiffness $\beta\gamma$ beyond the relaxed length $L_e$ but that will buckle and bear no load if shortened too much. We model the crosslinks as hard spheres of diameter $d_{HS}$ that provide the excluded volume. In the cytoskeletal instantiation this excluded volume represents the steric constraints from both the filaments and the binding proteins. Instead of following the filament degrees of freedom, we keep track of the motion of the crosslinks or the nodes of the network. Bonded node pairs interact with the cat's cradle type potential given by $\beta U(r)=\Theta(r-L_e)(1/2)\beta\gamma(r-L_e)^2+A\Theta(d_{HS}-r)$, where $\Theta(\cdot)$ is the Heaviside step function and $A\rightarrow\infty$ indicates hard-core repulsion. Elastic energy arises only when the contour length $r$ exceeds the relaxed length $L_e$.
The statistical architecture of the network is specified by the node density $\rho_0$ and network connectivity (with $P_c$ denoting the fraction of bonded neighbors). We assume the network architecture is quenched once initially assigned.

\begin{figure}[htb]
\centerline{\includegraphics[angle=0, scale=0.35]{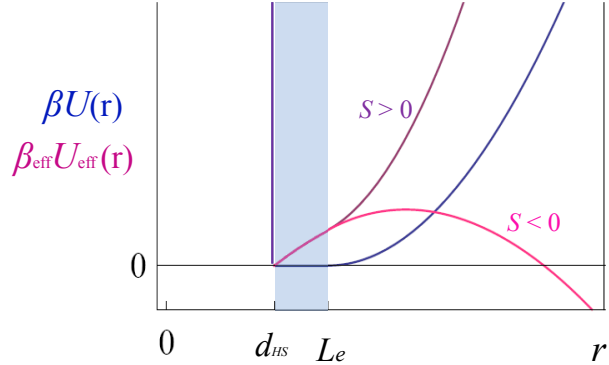}}
\caption{Schematic of model interactions. Blue line: $\beta U(r)$ for passive networks. Elastic stretching initiates at $r=L_e$. Purple ($s>0$) and Magenta ($s<0$) lines: $\beta_\textrm{eff}U_\textrm{eff}(r)$ for active networks. Motor action induces effective attraction even in the buckling regime (shaded area). Susceptible motors ($s>0$) enhance the long range attraction whereas load-resisting motors ($s<0$) may lead to long-range repulsion at high motor activity.}
\label{model_interaction}
\end{figure}

To deal with active network materials we have shown earlier \cite{effective interactions} that when agitated by spatially anti-correlated motor-driven events the cytoskeleton behaves as if it were at an effective equilibrium with a nontrivial effective temperature and modified interactions. The effective motor-driven pair interaction scaled by the effective temperature $T_\textrm{eff}$ ($=\beta_\textrm{eff}^{-1}$) is given by $\beta_\textrm{eff}U_\textrm{eff}(r)=\left[1+(s-1/2)\Delta\right]\beta U(r)+\Delta\ln r$, which depends on motor activity $\Delta$ and susceptibility $s$. A schematic illustration of the pair interaction in passive and motor-driven active networks is shown in Fig.~\ref{model_interaction}.
In contrast to the passive network (blue curve), motor action induces an effective attraction ($\Delta\ln r$ term) even in the buckling regime ($d_{HS}<r\leq L_e$, shaded area). Motor susceptibility modulates the long-range elastic interactions: Susceptible motors with $s>0$ (purple curve) enhance long-range attraction, whereas intense action of load-resisting motors with $s<0$ (magenta curve) may yield long-range repulsion.

Using the underlying parameters of the model and starting from an arbitrary initial configuration, one can perform molecular dynamics (MD) simulations to obtain representative fiducial structures for a given set of parameters ($\rho_0, P_c; L_e, \beta\gamma; \Delta, s$). These structures are vibrationally equilibrated but quenched as to density and connectivity. We fix the number density $\rho_0=1.2$ of a total of $N=256$ particles, setting the mean particle spacing $r_0$ as the length unit. Periodic boundary conditions are applied.

\begin{figure}[htb]
\centerline{\includegraphics[angle=0, scale=0.45]{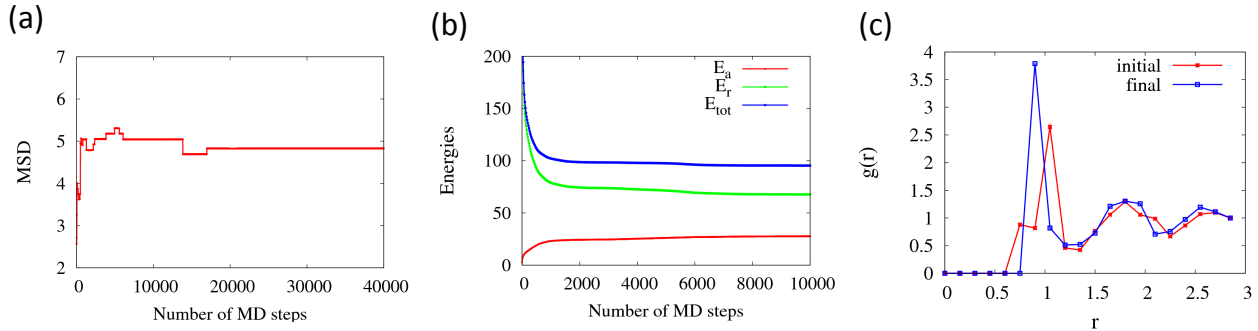}}
\caption{Monitoring the generation of fiducial structures. (a) Mean square displacement (MSD) versus number of MD steps. (b) Evolution of the potential energy. $E_a$ and $E_r$ denote for attractive and repulsive interactions, respectively, and $E_{tot}$ for their sum. (c) Radial distribution function g(r). Red: initial binary system; blue: final monodisperse structure.  Model parameters are $L_e=1.2, \beta\gamma=2.5, \eta=1$.}
\label{gen_fiducial}
\end{figure}

We start with a disordered configuration from a simulated Lennard-Jones binary mixture used for earlier studies of glassy dynamics \cite{magnetic analogy} but assume a single radius in the starting configuration for current purposes. We then evolve the configuration via MD steps until the mean square displacement (MSD) of the nodes with respect to their initial positions saturates to a plateau (see an example in Fig.~\ref{gen_fiducial}a).
During equilibration a series of jerks in the MSD arise in the absence of thermal noise (zero temperature assumed). These may be thought of as avalanches, corresponding to large-scale cooperative rearrangements.
During the MD evolution toward the steady state structure, as seen in Fig.~\ref{gen_fiducial}b, the repulsive interaction energy ($E_r$) decreases with time whereas the attractive interaction energy ($E_a$) due to bond stretching increases as structural rearrangements occur; the total potential energy ($E_{tot}$) decreases as time advances until an apparent steady state plateau is reached. The radial distribution functions $g(r)$ for the initial and final configurations are given in Fig.~\ref{gen_fiducial}c. These show how the original binary sphere system (red line) with two peaks in the innermost shell evolves into a monodisperse network structure (blue line) with a single peak in the first shell.

\section{Magnetic Analogies of the RFOT Theory}

A structural glass is statistically homogeneous but is specifically nonuniform at the microscopic level with a density $\rho(\vec{x})$ that is not translationally invariant. This is what gives rise to the glass's rigidity. Liquid state theory provides the free energy as a function of such a non-uniform density $F[\rho(\vec{x})]$. The one-to-one correspondence in the long time limit between force field and density profile has been elegantly established a l\`{a} the density functional theory \cite{DFT}. While the complete equilibrium free energy assumes the whole phase space can be sampled, close to the glass transition there occurs a trapping in locally metastable states which manifest themselves as an extensively large number of local free energy minima described by a configurational entropy $S_c=\ln N_{mss}$. Self-consistent phonon (SCP) theory directly addresses broken ergodicity by investigating the stability of a density wave frozen into aperiodic structures.

The random first order transition (RFOT) theory relates the activation energy of a rearranging unit (the ``droplet") to the microscopic intermolecular forces, and has successfully predicted many confirmed quantitative results for the dynamics of glass forming liquids \cite{Xia PNAS}.
The essence of the RFOT theory of escape from a local minimum is an analogy to a random field Ising magnet in a biasing field.
Stevenson \textit{et al.} \cite{magnetic analogy} have established an explicit mapping between a structural glass forming liquid and the corresponding short range disordered Ising magnet. This mapping yields a description of a structural glass in terms of discrete spin-like variables tied to the liquid structure equilibrated at one time.
The average field in the Ising magnet corresponds to the configurational entropy density $s_c$ for the model liquid, which provides an entropic driving force for the transitions among distinct configurational states. The smallness of $s_c$ indicates a paucity of accessible configurations, and thus signals a deep descent into the glassy regime. Fluctuations in configurational entropy, $\delta s_c$, on the other hand, are related to the excess heat capacity $\Delta C_p$ (according to Landau, $\delta s_c=\sqrt{\Delta C_p k_B/N_{corr}}$ where $N_{corr}$ is the volume within which the disorder is correlated) and thus measures the fragility of the liquid. The strength of the coupling $J_{ij}$ between neighboring Ising spins that point in opposite directions, is associated with the interaction energy between a low-overlap site and its high-overlap neighbor or vice versa.

In a three dimensional frozen aperiodic state, the density profile is well approximated by a sum of Gaussians centered around the fiducial lattice sites, $\rho(\vec{r})=\sum_i\rho_i(\vec{r})=\sum_i(\alpha_i/\pi)^{3/2}\exp[-\alpha_i(\vec{r}-\rif)^2]$, where \{$\alpha_i$\} represent the effective local spring constants that determine the mean square displacement from the fiducial sites $\{\rif\}$. For large values of $\{\alpha_i\}$ the particles are localized close to the fiducial locations. By using the independent oscillator approximation which yields site-wise decoupling of the particles, the free energy $\beta F_{glass}$ can be expressed as a sum of effective potentials between the interacting density clouds,
$\beta V_\textrm{eff}(|\rif-\rjf|;\alpha_j)=-\ln\int d\vec{r}_j\rho_j(\vec{r}_j)e^{-\frac{1}{2}\beta u(\rif-\vec{r}_j)}$. Within the Gaussian density ansatz one finds the free energy for the glassy state
\begin{equation}
\beta F_{glass}(\{\rif\}; \{\aiu\})=\sum_i\frac{3}{2}\ln\frac{\aiu\Lambda^2}{\pi}+\sum_{ij}\beta V_\textrm{eff}(|\rif-\rjf|;\aju).
\end{equation}

Near the low overlap state, the free energy follows from the equilibrium liquid equation of state (given by $Z_{EoS}(\eta)$) with corrections due to bonded constraints/interaction,
\begin{eqnarray}
\nonumber\beta F_{liq}(\{\rif\}; \{\aid\})&=&\sum_i\left(\frac{3}{2}\ln\frac{\aid\Lambda^2}{\pi e}-1\right)
+N\int_0^\eta[Z_{EoS}(\eta')-1]\frac{d\eta'}{\eta'}\\
&+&\sum_{ij}[\beta V_\textrm{eff}^{model}-\beta V_\textrm{eff}^{HS}](|\rif-\rjf|;\ajd).
\end{eqnarray}
Note the double summation in the last term only involves bonded pairs.

For any assignment of the \emph{discrete} values of $\{\aiu\}$ and $\{\aid\}$, the free energy of the model liquid is equivalent to a pairwise interacting model with spins located at the fiducial lattice sites, $\{\rif\}$,
\begin{equation}
\beta H=-\sum_i h_i(1-s_i)+\sum_{i<j}J_{ij}[s_i(1-s_j)+s_j(1-s_i)],
\end{equation}
where the spin $s_i=1$ corresponds to a large overlap site and $s_i=0$ a small overlap site. The average field is found from the bulk free energy difference between the states, $\sum_i h_i=\beta F_{glass}-\beta F_{liq}=Ns_c/k_B$, with a heterogeneous local field resulting from spatial variations of $\alpha$,
\begin{eqnarray}\label{local field}
\nonumber h_i&=&\frac{3}{2}\ln\frac{\aiu}{\aid}+\frac{5}{2}+\sum_j \beta V_\textrm{eff}(|\rif-\rjf|;\aju)
-\int_0^\eta[Z_{EoS}(\eta')-1]\frac{d\eta'}{\eta'}\\&-&\sum_{j}[\beta V_\textrm{eff}^{model}-\beta V_\textrm{eff}^{HS}](|\rif-\rjf|;\ajd).
\end{eqnarray}
Here the summation in the last term, i.e. the modification due to bonding, includes only bonded neighbors of the central node $i$.

The quenched interactions between neighboring low-overlap and high-overlap sites defined through the effective potential give the surface energies of the droplets within the RFOT picture and are explicitly
\begin{equation}\label{interaction}
J_{ij}=\beta V_2^\textrm{eff}(|\rif-\rjf|;\aid,\aju)+\beta V_2^\textrm{eff}(|\rif-\rjf|;\aiu,\ajd),
\end{equation}
where the pair interaction is given by
\begin{equation}\label{pair_interaction}
\beta V_2^\textrm{eff}(|\rif-\rjf|;\aid,\aju)=-\ln\int d\vec{r}_i d\vec{r}_j \rho^{\downarrow}_i(\vec{r}_i)\rho_j^{\uparrow}(\vec{r}_j)e^{-\frac{1}{2}\beta u(\vec{r}_i-\vec{r}_j)}.
\end{equation}

The statistics of the mapped out random fields (Eq.~\ref{local field}) and random interactions (Eq.~\ref{interaction}), which depend on the liquid structure, when combined with the renormalization group results for random magnets \cite{RG analysis} allow us to predict whether a thermodynamic phase transition would occur for the liquid analog when its mean field configurational entropy vanishes.

\section{Heterogeneous Self-Consistent Phonon (SCP) Method}

\begin{figure}[htb]
\centerline{\includegraphics[angle=0, scale=0.4]{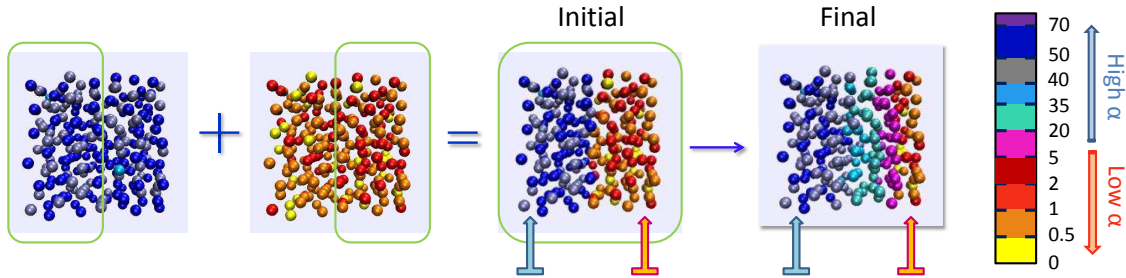}}
\caption{Illustration of interface formation. A quenched typical fiducial particle configuration with color-coded $\alpha$ values is shown. Color scale is given on the right hand side. Green frames indicate the creation of an initial sharp contact between two bulk stable phases. With the $\alpha$ values in the boundary layers being pinned (indicated by blue and red anchored arrows), the final $\alpha$ configuration develops a smooth interface, showing a gradual progression of phases.}
\label{interface_formation}
\end{figure}

The site-dependent Debye-Waller factors $\{\alpha_i\}$ can be determined using the heterogeneous SCP method applied to the frozen/quenched fiducial lattices generated by the molecular dynamics simulations.
The validity of the mapping relies on the self-consistency between the fiducial structure and the inter-particle potential, in other words, the one-to-one correspondence between the density profile and the force field, as well as a well separation of time scales between the vibrational motion about a fiducial configuration and the collective structural rearrangements representing hopping between different configurational states. Starting from \textit{uniform} low (of order $1$) and high (of order $100$) $\alpha$ values, we obtain two distinct sets of mechanically stable homogeneous solutions $\{\aid\}$ and $\{\aiu\}$, stabilized by bond stretching and hard-core repulsion, respectively. An example of the color-coded $\alpha$ configurations is shown in the left two panels of Fig.~\ref{interface_formation}. The modest spatial variation in $\alpha$ values that is seen reflects the disorder in quenched lattice structures and network connectivity.

The SCP method also allows us to study situations where the order parameters spatially vary across the sample. In particular, we can study the case where a smooth interface forms between liquid-like and solid-like bulk stable phases. Fig.~\ref{interface_formation} illustrates the formation of a broad interface (see configuration marked as ``final") from a mechanical relaxation (using SCP procedure) across an initially sharp interface (see configuration marked as ``initial") between distinct mobile (reddish) and localized (bluish) bulk states. The $\alpha$ values within the left and right boundary layers (indicated by anchored arrows) have been pinned during the relaxation to maintain the spatial gradient in $\alpha$.

\section{Main findings}

\subsection{Passive networks}

\subsubsection{Implications of the mapping}

$\diamondsuit$ \textit{Highly connected networks exhibit strong liquid behavior and descend deeper into the glassy regime}

\begin{figure}[htb]
\centerline{\includegraphics[angle=0, scale=0.55]{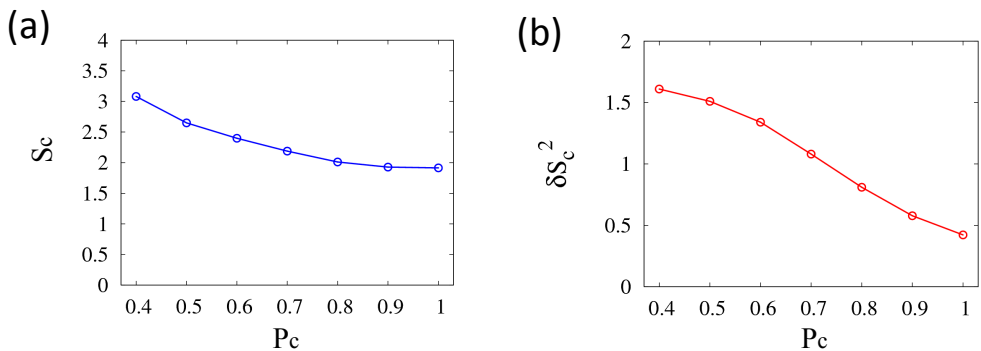}}
\caption{Highly connected networks descend deeper into the glassy regime and exhibit strong-liquid behavior. (a) The configurational entropy density $s_c$ vs connectivity $P_c$. (b) The fluctuations of configurational entropy density $\delta s_c^2$ vs connectivity $P_c$. Model parameters are $L_e=1.2, \beta\gamma=2.5, \eta=1$.}
\label{univ_Pc}
\end{figure}

We calculate the local fields $\{h_i\}$ defined by Eq.~\ref{local field} using the homogeneous solutions $\{\aiu\}$ and $\{\aid\}$ obtained by site-dependent SCP calculation.
According to the mapping, $\bar h$ is translated into the configurational entropy density, i.e., $\bar h=s_c/k_B$. In RFOT theory the configurational entropy parametrizes a liquid's descent into the glassy regime.
We show in Fig.~\ref{univ_Pc}(a) that increasing $P_c$ for a fixed density yields a smaller $s_c$, suggesting that crosslinking networks to a higher degree of connectivity leads them deeper into the glassy regime. This finding is consistent with earlier results using a statistically averaged RFOT-SCP theory for network glasses \cite{network glasses}. The result also agrees with the well-known experimental observation that the addition of nonbonding impurities into highly crosslinked network glasses lowers the glass transition temperature or raises the transition densities.
Physically, higher connectivity yields a much stronger localization even for the loosely tethered (liquid-like) state; the resulting change in the entropy cost to localize density waves dominates over the reduction of bonded interaction due to smaller fluctuations in motion, leading to a higher liquid-state free energy and hence a lower $s_c$ for the same configuration. This is because the glassy-state free energy itself is insensitive to $P_c$ since the caging of the motion arises primarily from steric constraints rather than bond stretching.

On the other hand, the fluctuations of the configurational entropy, $\delta s_c^2$, can be directly related to the excess heat capacity $\Delta C_p$. The mapping thus establishes the connection between a liquid's fragility and its degree of bonding: As seen in Fig.~\ref{univ_Pc}(b), higher $P_c$ yields smaller $\delta s_c^2$ and thus smaller $\Delta C_p$, indicating that liquids with a higher degree of bonding are stronger.
Again this coincides with experience for inorganic network glass formers, in addition, this trend is consistent with the experimental observation that F-actin disruption in the cytoskeleton substantially increases its fragility \cite{osmotically compressed cell}.

$\diamondsuit$ \textit{Quasi-universality with respect to network connectivity in deeply glassy regime}

\begin{figure}[htb]
\centerline{\includegraphics[angle=0, scale=0.6]{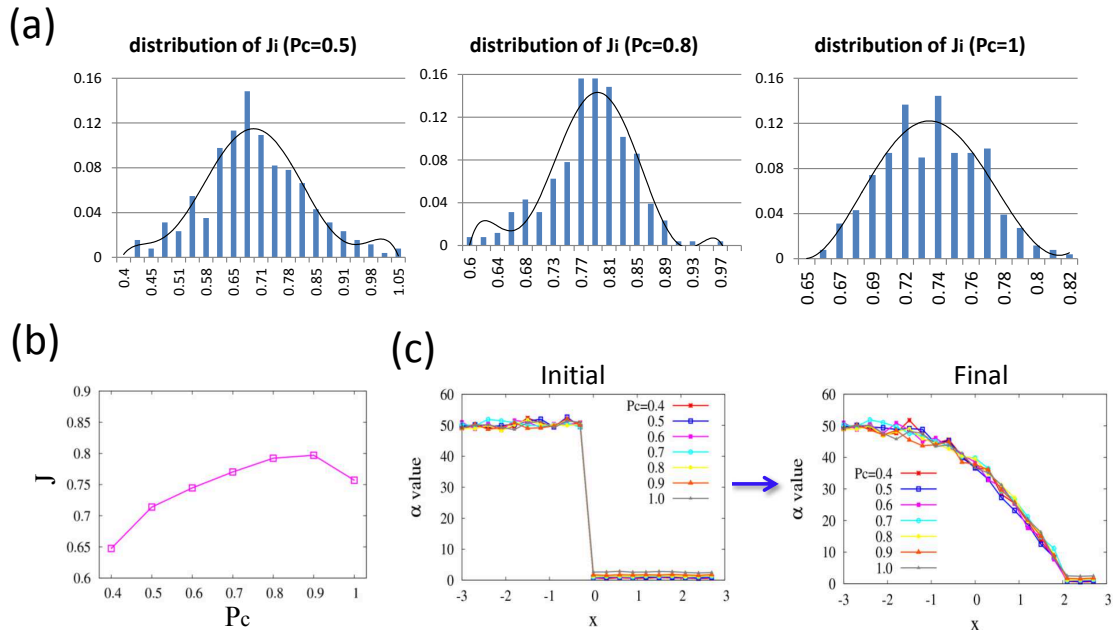}}
\caption{Quasi-universality of the surface energy and the interface width with respect to network connectivity. (a) Distribution of the calculated local interactions $\{J_i\}$ of the analogous magnets. Left to right: $P_c=0.5, 0.8$ and $1$. (b) $J$ vs $P_c$. Typical surface tension $J$ ($\equiv(1/N)\sum_i J_i$) is close to the RFOT estimate $J_{RFOT}\simeq0.58$. (c) Initial and final $\alpha$ profiles at various $P_c$. A broad interface forms between localized and mobile bulk states. Model parameters are $L_e=1.2, \beta\gamma=2.5, \eta=1$.}
\label{univ_J_width}
\end{figure}

We have examined the statistics of the mapped out random interactions according to Eq.~\ref{interaction}. Fig.~\ref{univ_J_width}(a) displays the distributions, at various levels of crosslinking, $P_c$, of the calculated interactions as the free energy per neighbor, $J_i\equiv\frac{1}{z_i}\sum_j J_{ij}$, where $z_i$ is the coordination number of particle $i$. The typical interaction $J\equiv\frac{1}{N}\sum_i J_i$ is related to the mismatch free energy penalty in RFOT theory for a particle at a flat interface between regions of high and low overlap.
The Xia-Wolynes estimate \cite{Xia PNAS} for surface tension gives $J_{RFOT}=\frac{3}{4}k_B T\ln\frac{1}{d_L^2\pi e}/n_{bb}\simeq0.58$,
where $d_L=(1/10)r_0$ is the Lindemann length and $n_{bb}=3.2$ is the typical number of bonds broken by the interface.
As shown in Fig.~\ref{univ_J_width}(b), the calculated $J$ for single-particle droplets (low-$\alpha$ sites against a high-$\alpha$ glassy background) is insensitive to $P_c$ and has a similar value to the Xia-Wolynes estimate.
The absence of mechanical relaxation across the sharp interface would explain the modest overestimation of the energetic penalty.
Fig.~\ref{univ_J_width}(c) demonstrates the quasi-universal profile of the broad interface with respect to $P_c$ if mechanical relaxation between the pinned boundaries is allowed, such that a gradual progression of phases lowers the energetic cost of forming an interface.

$\diamondsuit$ \textit{Locating analogous magnets on the Renormalization Group phase diagram}

\begin{figure}[htb]
\centerline{\includegraphics[angle=0, scale=0.35]{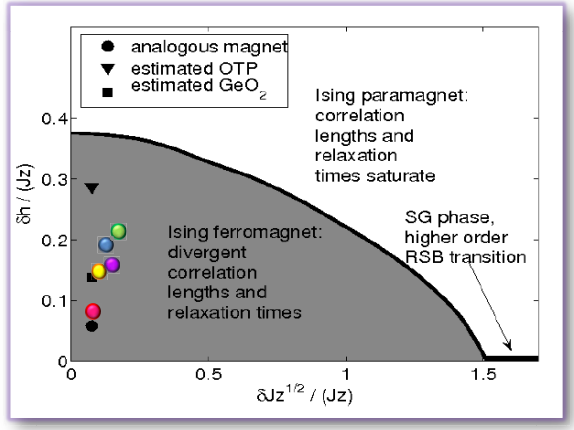}}
\caption{Locating analogous magnets on the RG phase diagram for a random field Ising model at zero average field (adapted from Fig.~3 in Ref.~[3]). The colored symbols correspond to model networks with $L_e=1.8, P_c=0.5$ at $\beta\gamma=5$ (green) and $20$ (purple), and to those with $L_e=1.2, \beta\gamma=2.5$ at $Pc=0.5$ (blue), $0.8$ (yellow) and $1$ (red). The analogous magnets are located close to the strong glass forming liquid $\textrm{GeO}_2$ (square mark).}
\label{ana_magnets}
\end{figure}

Analysis of the statistics of the calculated random fields and interactions also allows us to locate the analogous magnets on the random field Ising magnet phase diagram determined by renormalization group analysis at zero average field (Fig.~\ref{ana_magnets}).
Both the field fluctuations, $\delta h$, and the fluctuations of the interaction strength, $\delta Jz^{1/2}$, are normalized by the total interaction energy per site $\bar Jz$. We use the $J_i$ distributions determined for single-particle droplets.


In Fig.~\ref{ana_magnets} the colored symbols indicate estimates of where the analogous magnets for model networks with various nonlinearity parameters and connectivity would fall on the magnetic phase diagram (see figure caption for detailed parameters). We find that
a highly-connected (large $P_c$) tense (short $L_e$) network consisting of stiff bonds (large $\beta\gamma$) is located deeply inside the ferromagnetic phase region, implying that the model network would experience a true phase transition to a state with infinite correlation length and divergent relaxation time when the mean field configurational entropy vanishes.
The associated ideal glass transition may underlie the observed glassy dynamics of living cells \cite{cytoskeletal slow dynamics}.


Moreover, the analogous magnets are located close to the strong glass forming liquid $\textrm{GeO}_2$.
Recent experiments on osmotically compressed cells by Weitz group \cite{osmotically compressed cell} suggest that cells under compressive stress indeed behave as strong colloidal glass formers, based on an observed Arrhenius-type exponential growth of viscosity for cells with increasing volume fraction, as well as their lower fragility compared to hard spheres.

\subsubsection{Variation of surface tension and interface width with reducing packing fraction: existence of a spinodal point}

$\diamondsuit$ \textit{The surface free energy density $\sigma$}

\begin{figure}[htb]
\centerline{\includegraphics[angle=0, scale=0.6]{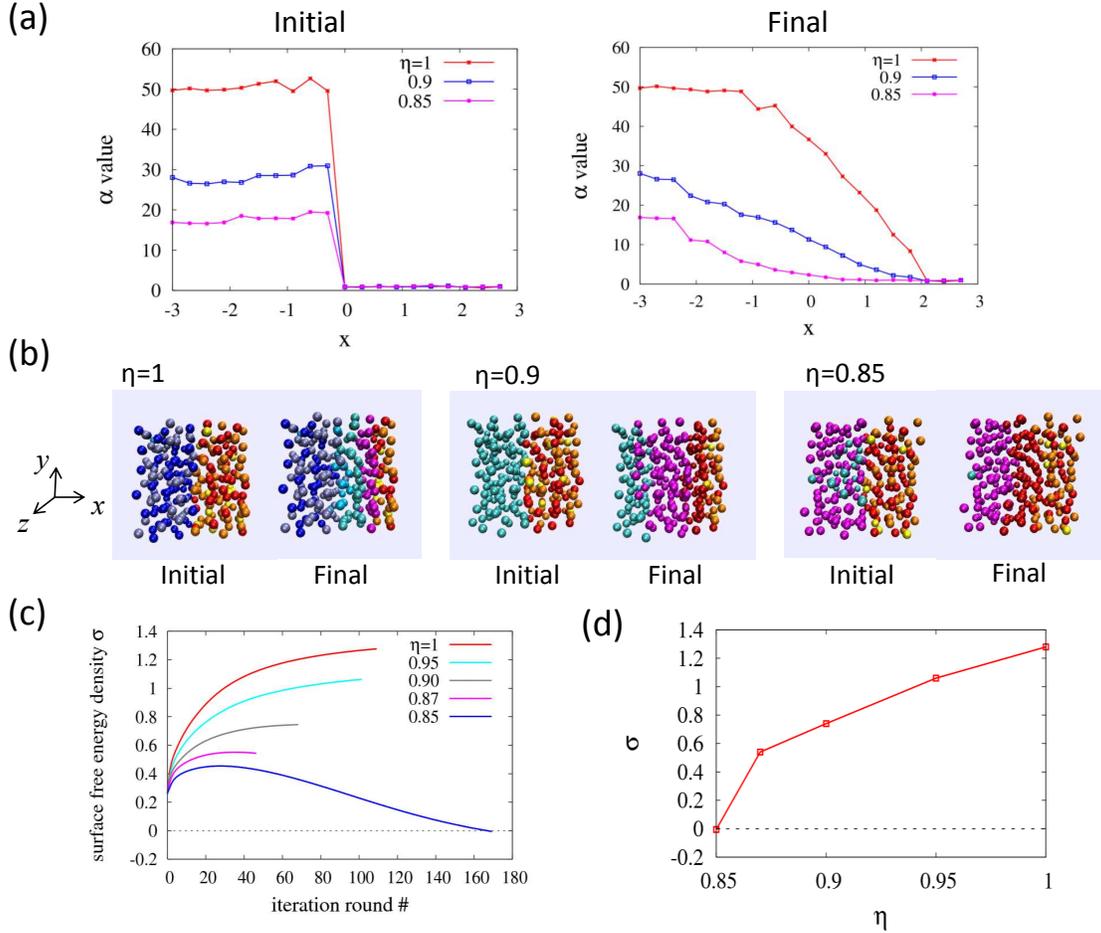}}
\caption{The variation of interface energetics and structure with reducing packing fraction $\eta$. (a) Initial and final $\alpha$ profiles along x axis at various packing fractions. Homogeneous high $\alpha$ and low $\alpha$ solutions get closer as $\eta$ decreases. (b) Corresponding color-coded $\alpha$ configurations. Same color scheme as in Fig.~\ref{interface_formation}. (c) Surface tension versus iteration round number. Final surface tension almost vanishes as $\eta$ approaches $\eta_A\simeq0.85$. (d) Surface tension versus packing fraction. The model parameters are $L_e=1.2, \beta\gamma=2.5, P_c=0.5$.}
\label{sigma_eta}
\end{figure}

In the same way that Cammarota \textit{et al.} \cite{surface tension} computed the surface energy for hybrid inherent structures at zero temperature, we can define the surface free energy ($F_s$) as the free energy cost of forming a stable interface between mobile and localized regions, with respect to the total bulk energies of the two regions if they were to exist separately having homogeneous low and high $\alpha$ values respectively.
Mathematically, we define $F_s\equiv F_{tot}-(F^0_L+F^0_R)$, where $F^0_{L(R)}=\sum_{i\in L(R)}\left[(3/2)\ln(\alpha_i^{\uparrow(\downarrow)}/\pi)+\sum_{j\in L(R)}\beta V_{eff}(|\rif-\rif|;\alpha_j^{\uparrow(\downarrow)})\right]$ represents the bulk energy of the left (right) region with homogeneous $\{\alpha_i^{\uparrow}\}$ ($\{\alpha_i^{\downarrow}\}$) solutions (see the initial configurations in Fig.~\ref{sigma_eta}b), and $F_{tot}$ denotes the total free energy of the stable inhomogeneous $\alpha$ configuration in which there is a smooth interface interpolating between the pinned boundary layers (see the final configurations in Fig.~\ref{sigma_eta}b).
At finite temperatures, the stable $\alpha$ configurations represent a compromise between the cost of localizing a particle and the free energy gain realized by particles being able to avoid each other once localized.

To investigate the possible existence of a spinodal point, we studied what happens as we reduce the packing fraction by shrinking the particles while maintaining their number density. Another extreme case of great interest is the jamming limit where particles run into each other and get firmly stuck. In this case the reconfiguration barrier becomes too high for remodeling to be relevant on laboratory time scales. Thus in the jamming regime mean-field theories should be sufficient for many practical purposes.

We show in Fig.~\ref{sigma_eta} the initial and final $\alpha$ profiles at various packing fractions (panel a) and the corresponding color-coded $\alpha$ configurations (panel b).
Panel d displays the surface free energy density (or surface tension) $\sigma$ defined as $\sigma\equiv F_{s}/N$, obtained by SCP calculations (i.e. final values in panel c), versus the packing fraction $\eta$. It is clearly seen that surface tension decreases with reducing packing fraction and ultimately vanishes at $\eta\simeq0.85$. Concomitantly we find that, for $\eta<0.85$, high-$\alpha$ solutions corresponding to localized amorphous packing become mechanically unstable to even small thermal fluctuations. This limit of stability for an amorphous system is comparable to the Lindemann criterion for melting of crystalline solids. Therefore, surface tension and mechanical stability of amorphous packing seem to vanish at a similar $\eta$, the dynamic transition packing fraction $\eta_A$, which can therefore be identified as a spinodal point.
Consistent with this interpretation we find, homogeneous high and low $\alpha$ solutions tend to merge as $\eta$ decreases toward $\eta_A$.

$\diamondsuit$ \textit{Quantifying the interface width $\xi$}

\begin{figure}[htb]
\centerline{\includegraphics[angle=0, scale=0.56]{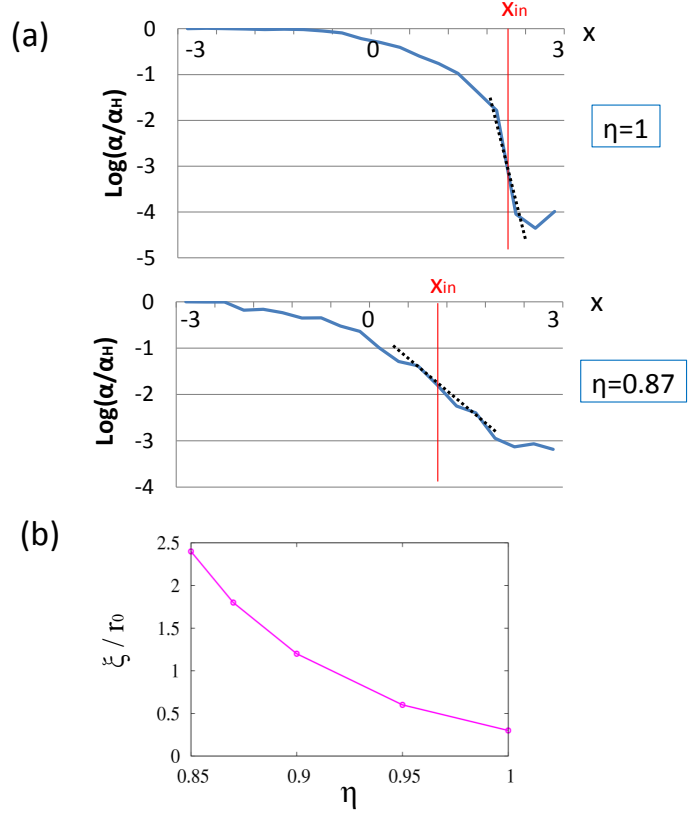}}
\caption{The variation of interface width ($\xi$) with reducing packing fraction $\eta$. (a) Quantification of the interface width at $\eta=1$ (upper) and $\eta=0.87$ (lower). $x_{in}$ indicates the location of interface where most significant variation in $\alpha$ occurs. The black dotted line represents a linear fit to the $\log(\alpha/\alpha_H)$ versus $x$ curve at $x_{in}$. (b) The interface width $\xi$ in units of inter-particle spacing $r_0$ versus packing fraction. The interface broadens as $\eta$ decreases toward $\eta_A$. Model parameters are $L_e=1.2, \beta\gamma=2.5, P_c=0.5$.}
\label{interface_width_eta}
\end{figure}

In Fig.~\ref{interface_width_eta}(a) we plot $\log(\alpha/\alpha_H)$ versus $x$ along which spatial variation of $\alpha$ occurs. We denote by $\alpha_H$ ($\alpha_L$) the high (low) $\alpha$ value at the pinned boundary to the left (right).
Much above $\eta_A$ (see $\eta=1$ case), the interface is quite sharp; in a single layer $\alpha$ changes from near $\alpha_H$ to near $\alpha_L$. Close to $\eta_A$ (see $\eta=0.87$ case), the transition becomes smoother with $\alpha$ slowly varying over several atomic layers. In both cases, there arises a surface tension $\sigma$ reflecting the deviation of $\alpha$ in the interfacial layers from the $\alpha$ for either of the bulk free energy minima.

To quantify the interface width,
we locate the interface at the position $x_{in}$ (red vertical lines in Fig.~\ref{interface_width_eta}a) where the most significant spatial variation of $\alpha$ takes place.
We then make a linear fit (black dotted lines) at $x_{in}$ and define the interface width $\xi$ as the inverse slope of the curve $\log(\alpha/\alpha_H)$ versus $x$, i.e. $\frac{1}{(\xi/r_0)}\equiv\Big|\frac{d\log(\alpha/\alpha_H)}{d(x/r_0)}\Big|_{x=x_{in}}$.
As shown in Fig.~\ref{interface_width_eta}(b), the interface gradually spreads out as the packing fraction decreases, corresponding to a lower free energy cost for a smooth spatially varying order parameter. At the spinodal point, a vanishing surface tension indicates the formation of a sufficiently smooth interface between bulk stable phases no longer costs any energy.

\begin{figure}[htb]
\centerline{\includegraphics[angle=0, scale=0.65]{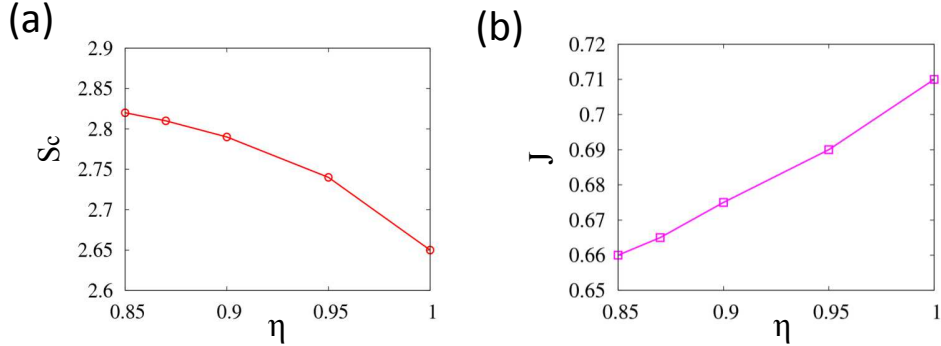}}
\caption{The variation of configurational entropy $s_c$ (a) and surface tension $J$ (b) with reducing packing fraction $\eta$. The model parameters are $L_e=1.2, \beta\gamma=2.5, P_c=0.5$.}
\label{Sc_J_eta}
\end{figure}

Fig.~\ref{Sc_J_eta} demonstrates how the glassy parameters vary with the packing fraction. As $\eta$ is reduced, configurational entropy density $s_c$ increases (panel a) whereas surface tension $J$ for single-particle droplets moderately decreases (panel b). But $J$ does not vanish when $\sigma$ does. Instead, as $\eta$ approaches the spinodal point ($\eta_A=0.85$), $J$ remains rather close to the Xia-Wolynes estimate.
Moreover, the explicit mapping gives an entropy at crossover $s_c(\eta_A)\simeq2.8k_B$. This is approximately double the estimate given by the string transition theory \cite{string_transition}; the latter predicts barrierless reconfiguration events occur at a critical configurational entropy $s_c^{string}=1.13k_B$. It is also quite a bit larger than the estimate based on percolation clusters $s_c^{perc}=1.28k_B$. These results are consistent with the observation that activated dynamics and mode coupling effects coexist at the empirical crossover temperature \cite{Sastry, network glasses, facilitation}.

\subsection{Active networks: dependence of glassy characteristics on motor properties}

While passive glassy systems remain trapped in micro-configurations for incredibly long times because only thermal energy is available to allow the system to evolve,
ATP-dependent rearrangements of active cytoskeletal networks modify the micro-configurations themselves and provide
an alternate means of exploring new configurations. ATP-hydrolysis-powered motor action can resolve steric constraints and drive structural rearrangements.

\subsubsection{Effect of motor action}

\begin{figure}[htb]
\centerline{\includegraphics[angle=0, scale=0.58]{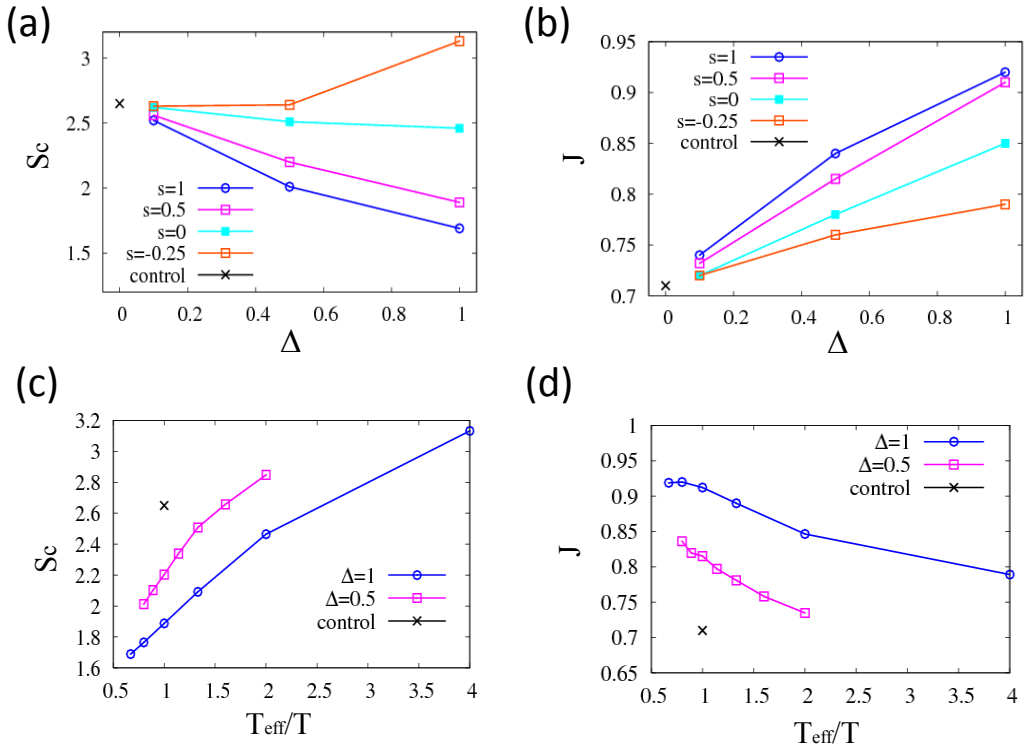}}
\caption{Glassy characteristics for active networks. The dependence of configurational entropy density $s_c$ (panels a and c) and surface tension $J$ (panels b and d) on motor activity $\Delta$, susceptibility $s$ and effective temperature $T_\textrm{eff}$ is shown. We also indicate by black crosses the corresponding quantities for passive networks with the same underlying architecture (at $\Delta=0$ and thus $T_\textrm{eff}=T$) for comparison. Model parameters are $L_e=1.2, \beta\gamma=2.5, P_c=0.5, \eta=1$.}
\label{active_glassy}
\end{figure}

We have shown earlier \cite{effective interactions} that a model cytoskeleton driven by small-step motor kicks can be described by an effective temperature $T_\textrm{eff}$ that depends on motor activity $\Delta$ and motor susceptibility $s$ through the relation $T_\textrm{eff}/T=\left[1+(s-1/2)\Delta\right]^{-1}$. Using this explicit mapping, we show in Fig.~\ref{active_glassy} that for a given $\Delta$, as $s$ decreases and thus $T_\textrm{eff}$ rises, configurational entropy $s_c$ increases (panels a and c) whereas the surface tension $J$ decreases (panels b and d), indicating that the higher $T_\textrm{eff}$ induced by increasingly more load-resisting motors at a given activity promotes reconfiguration and reduces the mismatch free energy penalty.
On the other hand, as seen in panel (b), for any given value of $s$, higher motor activity leads to a larger surface tension $J$. This is because intense motor action enhances the effective attraction in the buckling regime, regardless of motor susceptibility.
More interestingly, as shown in panel (a), while $s_c$ decreases with increasing $\Delta$ for susceptible motors ($s>0$), load-resisting motors ($s=-0.25$) might resolve constraints and promote structural remodeling, thus expanding the accessible configurational state space as motor activity increases, i.e.~ there is a larger $s_c$ as $\Delta$ increases.

As a comparison, the black crosses in the figures show the values of $s_c$ and $J$ in the absence of motor processes, i.e. at $\Delta=0$ and thus $T_\textrm{eff}=T$. These correspond with what should be found after complete ATP depletion in experiments. We might say after ``rigor mortis has set in". The convergence of different curves to the passive values as $\Delta\rightarrow0$ is apparent (see panels a and b).
While panels (a) and (b) disentangle the effect of $\Delta$ and $s$ on glassy dynamics, panels (c) and (d) demonstrate the use of $T_\textrm{eff}$ in describing the non-equilibrium behavior driven by small-step motors: Raising $T_\textrm{eff}$ allows the system to explore a larger portion of the configuration space (larger $s_c$) and reduces the coupling strength (lower $J$).

\subsubsection{Estimate of the activation barrier}

\begin{figure}[htb]
\centerline{\includegraphics[angle=0, scale=0.75]{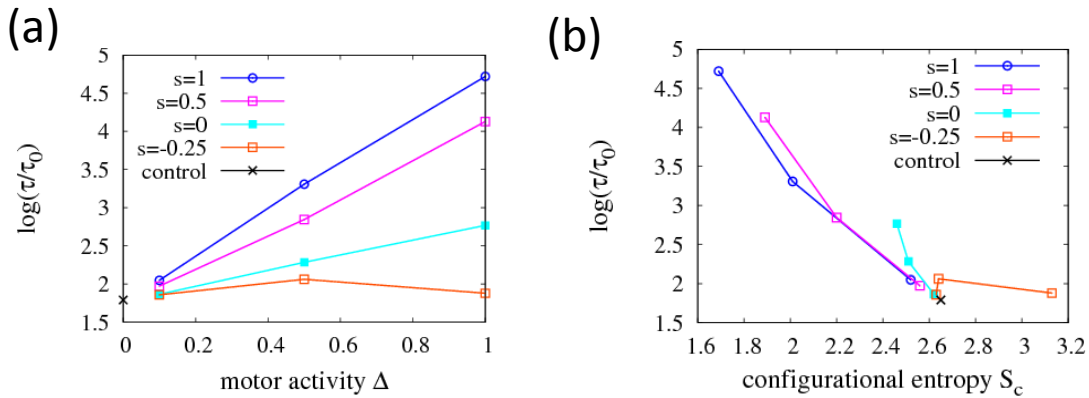}}
\caption{The RFOT estimate of the magnitude of activation barrier. We plot barrier height $\log(\tau/\tau_0)$ versus (a) motor activity $\Delta$ and (b) configurational entropy $s_c$, for a series of motor susceptibility $s$. ``Control" gives the passive value at $\Delta=0$ for comparison. Model parameters are $L_e=1.2, \beta\gamma=2.5, P_c=0.5, \eta=1$.}
\label{barrier}
\end{figure}

RFOT theory allows us to estimate the magnitude of activation barriers. Using the expression from RFOT theory that includes the interface disorder wetting effect \cite{Xia PNAS} we can write
\begin{equation}
\log(\tau/\tau_0)=\beta_\textrm{eff}\Delta F^{\ddag}=3\pi\frac{J^2}{s_c}.
\end{equation}
Fig.~\ref{barrier} shows the dependence of the resulting activation barrier height $\beta_\textrm{eff}\Delta F^{\ddag}$ on motor activity $\Delta$ (panel a) and configurational entropy $s_c$ (panel b) for a series of motor susceptibility $s$. For susceptible motors with $s>0$, more intense motor action raises the reconfiguration barrier. Load-resisting motors ($s=-0.25$) at sufficiently high activity, however, actually facilitate structural rearrangements thus lowering the barrier (see lowest curve in panel a). An inverse dependence of barrier height on the configurational entropy remains valid (see panel b).

To translate the barrier height into relaxation times $\tau$ in the laboratory, we need to estimate the bare relaxation time $\tau_0$.  $\tau_0=d_L^2/D_\textrm{eff}$ is the time to diffuse a Lindemann length $d_L$, which is related to the localization strength $\alpha_H$ of confined motion through $d_L^2=1/\alpha_H$. The effective diffusion constant given by $D_\textrm{eff}=k_BT(1+\Delta/3)/6\pi\eta_0r$ depends inversely on the viscosity $\eta_0$ of the medium in which the network is immersed as well as the effective radius $r$ of the crosslinks/nodes of the network.
The drastic difference in relaxation time between molecular glass formers and colloidal glass formers lies in the big difference in their $\tau_0$.

\begin{figure}[htb]
\centerline{\includegraphics[angle=-90,scale=0.28]{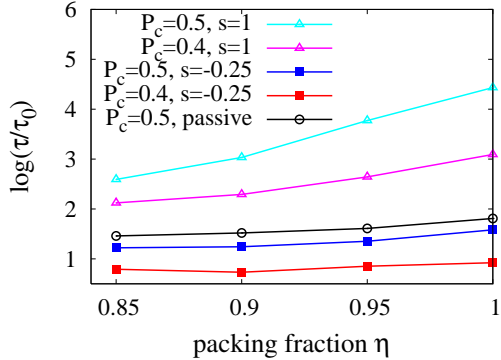}}
\caption{Dependence of reconfiguration barrier on motor susceptibility and network connectivity. We plot barrier height $\log(\tau/\tau_0)$ versus packing fraction $\eta$ for susceptible motors ($s=1$, triangle) and load-resisting motors ($s=-0.25$, square) at different network connectivity $P_c$. Passive curve (with black circles) corresponds to ATP-depletion in experiments. Model parameters are $L_e=1.2, \beta\gamma=2.5, \Delta=1$.}
\label{ATP_Pc_dep}
\end{figure}

The Weitz group \cite{osmotically compressed cell} has recently shown that cells under compressive stress behave as strong colloidal glass formers; ATP depletion raises the reconfiguration barrier, whereas F-actin disruption accelerates remodeling. Our theory predicts a trend consistent with these observations for the dependence of glassy dynamics on the nonequilibrium processes and network connectivity.
Fig.~\ref{ATP_Pc_dep} demonstrates how the activation barrier height varies with packing fraction at various network connectivities and motor susceptibilities. The passive case (black, circle) without motor action corresponds to ATP depletion in experiment. Compared to the active network at the same connectivity ($P_c=0.5$) yet driven by load-resisting motors ($s=-0.25$; blue square), the lack of motor processes in the passive case indeed raises the activation barrier. Such motor-facilitated remodeling is also consistent with active fluidization observed in polymer networks where myosin-II motors enhance longitudinal filament motion \cite{active fluidization}.
On the other hand, a reduction in connectivity ($P_c=0.4, s=-0.25$; red square) mimicking F-actin disruption leads to a faster relaxation, since fewer bond constraints yield weaker interactions and larger configurational entropies and hence a lower reconfiguration barrier.

In contrast to the load-resisting motors, at a given connectivity, susceptible motors ($s=1$, lines with triangular marks) raise the activation barrier and suppress remodeling. This conflicting behavior suggests a way to determine the sign of the effective motor susceptibility in living cells. The experimental observation that active processes promote reconfiguration for a broad range of cell types \cite{osmotically compressed cell} thus suggests molecular motors in the cytoskeleton tend to work against mechanical load ($s<0$) rather than move energetically downhill ($s>0$).
As we have shown earlier, load-resisting motor action is also essential for macroscopic active contractility in actomyosin networks \cite{contractility}.
Finally, in all the cases, reconfiguration slows down (i.e.~$\tau$ increases) as the volume fraction $\eta$ increases, as seen experimentally \cite{osmotically compressed cell}.

\section{Summary}

In this work, we have developed a theoretical framework to study the glassy dynamical behavior of passive and active network materials. To illustrate the concepts, we have built a general microscopic network model that incorporates both the nonlinear elasticity of individual filaments and the steric constraints due to crowding. By treating the network materials as structural glass forming liquids, we draw an explicit analogy to disordered short-range Ising magnets. This magnetic analogy, when combined with renormalization group phase diagrams for a random-field Ising model, allows us to predict whether a thermodynamic phase transition underlies the observed glassy dynamics in network materials. The calculations also locate the model networks close to strong glass formers, as found experimentally.

Moreover, this explicit mapping provides a microscopic route to compute the mismatch surface tension of RFOT theory and the configurational entropy and thus lends insight on how the network architecture and motor properties influence the activation dynamics associated with structural rearrangements. Consistent with experimental observations in living cells, we find that intense action of load-resisting motors allows the systems to explore a larger portion of the configurational space and thus facilitates more efficient remodeling.
A heterogeneous self-consistent phonon approximation enables us to study spatially varying order parameters with a smooth gradient across the system. This method proves useful in identifying a spinodal point where formation of a sufficiently smooth interface between localized and freely moving particles no longer costs any energy and activated dynamics crosses over to collisional transport.
Finally, near-universality of the interaction per molecular unit based on universality of Lindemann ratio of all glass formers made of spherical particles still proves to be a good approximation in deeply glassy regime (i.e. at sufficiently high packing fraction) for both passive and active network materials.

\begin{acknowledgments}
We gratefully acknowledge the financial support from the Center for Theoretical Biological Physics
sponsored by the National Science Foundation via Grant PHY-0822283 and the additional support from the Bullard-Welch Chair at Rice University.
\end{acknowledgments}


\end{document}